\documentclass[submission,copyright,creativecommons]{eptcs}
\providecommand{\event}{PLACES 2010} 
\usepackage{breakurl}             

\usepackage{amsmath,amssymb}
\usepackage{graphicx} 
\usepackage{xspace} 
\usepackage[english]{babel} 
\usepackage[inference,ligature]{semantic}
\usepackage{subfigure}
\usepackage{times}
\usepackage{enumerate}
\usepackage{wrapfig}

\title{Declarative Event-Based Workflow as \\Distributed Dynamic Condition Response Graphs}
\author{Thomas T. Hildebrandt \quad  \quad Raghava Rao Mukkamala\\ 
\{hilde,rao\}@itu.dk\\
IT University of Copenhagen \\ Programming, Logic and Semantics Group\\ Rued Langgaards Vej 7, DK-2300 Copenhagen S, Denmark\\
}

\def\titlerunning{Declarative Event-Based Workflow as Distributed DCR Graphs}
\def\authorrunning{Hildebrandt and Mukkamala}

\newcommand{\conditionrel}{\ensuremath{\mathrel{\rightarrow\!\!\bullet}}}

\newcommand{\responserel}{\ensuremath{\mathrel{\bullet\!\!\rightarrow}}}
\newcommand{\condresponserel}{\ensuremath{\mathrel{\bullet\!\!\rightarrow\!\!\bullet}}}
\newcommand{\includerel}{\ensuremath{\mathrel{\rightarrow\!\!+}}}
\newcommand{\inex}{\ensuremath{\pm}}
\newcommand{\excluderel}{\ensuremath{\mathrel{\rightarrow\!\!\%}}}
\newcommand{\as}{\ensuremath{\mathop{\mathsf{as}}}}

\newcommand{\dcrg}{\ensuremath{\mathsf{DCR\, Graphs}}}

\newcommand{\roles}{\ensuremath{\mathsf{Roles}}}
\newcommand{\power}[1]{\ensuremath{{\cal{P}}(#1)}}
\newcommand{\agents}{\ensuremath{\mathsf{P}}}
\newcommand{\events}{\ensuremath{\mathsf{E}}}

\newcommand{\actions}{\ensuremath{\mathsf{Act}}}
\newcommand{\modelshort}{\dcrg}
\newcommand{\modelname}{dynamic condition response graph}

\newcommand{\modelnamebb}{Dynamic Condition Response Graphs}

\newcommand{\conreseventstructure}{condition response event structure}
\newcommand{\cres}{\ensuremath{\mathsf{CRES}}}
\newcommand{\es}{\ensuremath{\mathsf{ES}}}
\newcommand{\marking}{\ensuremath{\mathsf{M}}}

\newcommand{\markingset}{\ensuremath{\mathcal{M}}}

\newcommand{\executed}{\ensuremath{\mathsf{Ex}}}
\newcommand{\included}{\ensuremath{\mathsf{In}}}
\newcommand{\responses}{\ensuremath{\mathsf{Re}}}

\newtheorem{definition}{Definition}
\newtheorem{theorem}{Theorem}
\newtheorem{proposition}{Proposition}

\begin{document}
\maketitle

\begin{abstract}
We present  \emph{\modelnamebb} (\modelshort)  as a declarative, event-based process model inspired by the workflow language employed by our industrial partner and conservatively generalizing prime event structures. 
A dynamic condition response graph is a directed graph with nodes representing the events that can happen and arrows representing four relations between events: condition, response, include, and exclude. 
Distributed \modelshort{} is then obtained by assigning roles to events and principals. We 
give a graphical notation inspired by related work by van der Aalst et al. 
We
exemplify the use of distributed 
\modelshort{} on a simple 
workflow taken from a field study at a Danish hospital, pointing out their flexibility compared to imperative workflow models. Finally we provide a mapping from \modelshort{} to B\"uchi-automata.
\end{abstract}

\section{Introduction}
\label{sect:introduction}
A key difference between declarative and imperative process languages is that the  control flow for the first kind is defined \emph{implicitly} as a set of constraints or rules, and for the latter is defined \emph{explicitly}, e.g. as a flow diagram or a sequence of state changing commands.

There is a long tradition for using declarative logic based languages to schedule transactions in the database community, see e.g.~\cite{logicbasedactivedatabases}.
Several researchers have noted~\cite{Davulcu98logicbased,Senkul02alogical, eventalgebra, agentcoorddbsystems,DeclWFAalstPS09,ConDec:2006, Pesic2008} that it could be an advantage to use a declarative approach
to achieve more flexible process descriptions in other areas, in particular for the specification of case management workflow and ad hoc business processes. The increased flexibility is obtained in two ways: Firstly, since it is often complex to explicitly model all possible ways of fulfilling the requirements of a workflow, imperative descriptions easily lead to over-constrained control flows.
In the declarative approach any execution fulfilling the constraints of the workflow is allowed, thereby leaving maximal flexibility in the execution.
Secondly, adding a new constraint to an imperative process description often requires that the process code is completely rewritten, while the declarative approach just requires the extra constraint to be added. In other words, declarative models provide flexibility for the execution at run time and with respect to changes to the process.

As a simple motivating example, consider a hospital workflow extracted from a real-life study of paper-based oncology workflow at danish hospitals~\cite{prohealth2008,DDBP:2008}. As a start, we assume two events, \emph{prescribe} and \emph{sign}, representing a doctor adding a prescription of medicine to the patient record and signing it respectively. We assume the constraints stating that  the doctor must sign after having added a prescription of medicine to the patient record and not to sign an empty record. A naive imperative process description may simply put the two actions in sequence, \emph{prescribe;sign}, which allows the doctor first to prescribe medicine and then sign the record. In this way the possibilities of adding several prescriptions before or after signing and signing multiple times are lost, even if they are perfectly legal according to the constraints. The most general imperative description should start with the prescribe event, followed by a loop allowing either sign or prescribe events and only allow termination after a sign event. If the execution continues forever, it must be enforced that every prescription is eventually followed by a sign event.

With respect to the second type of flexibility, consider adding a new event \emph{give}, representing a nurse giving the medicine to the patient, and the rule that a nurse must give medicine to the patient if it is prescribed by the doctor, but not before it has been signed.
 For the most general imperative description we should add the ability to execute the \emph{give} event within the loop after the first \emph{sign} event and not allow to terminate the flow if we have had a \emph{prescribe} event without a subsequent \emph{give} event. So, we have to change the code of the loop as well as the condition for exiting it.
 
\begin{wrapfigure}{R}{0.40\textwidth}
  \vspace{-20pt}
  \begin{center}
    \includegraphics[width=0.40\textwidth]{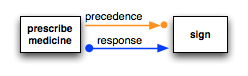}
  \end{center}
  \vspace{-20pt}
  \caption{Graphical notation proposed in~\cite{ConDec:2006, Pesic2008}.}
  \label{fig:prescribesign}  
  \vspace{-10pt}
\end{wrapfigure}

In~\cite{ConDec:2006, Pesic2008}, van der Aalst and Pesic propose to use Linear-time Temporal Logic (LTL) as a declarative language for describing the constraints of the workflow. LTL allows for describing a rich set of constraints on the execution flow. In particular, the first example workflow above is expressed as $(\mathbf{F}\text{Prescribe}\implies\neg\text{Sign}\mathrel{\mathbf{U}}\text{Prescribe})\wedge (\mathbf{G}(\text{Prescribe}\implies\mathbf{F}\text{Sign}))$,  in words: "(Future Prescribe implies (not Sign Until Prescribe)) and (Globally (Prescribe implies Future Sign))". The expression becomes slightly more readable if the past modality is used: $(\mathbf{G}(\text{Sign}\implies\mathbf{P}\text{Prescribe})\wedge (\mathbf{G}(\text{Prescribe}\implies\mathbf{F}\text{Sign}))$,  in words: "(Globally (Sign implies Past Prescribe)) and (Globally (Prescribe implies Future Sign))".
Since the notation of LTL is likely to be too difficult to use directly by the end user it is suggested to use a graphical notation for common patterns of temporal constraints which are then compiled to LTL. The example is then a combination of a \emph{preceedence} pattern, $(\mathbf{G}(\text{Sign}\implies\mathbf{P}\text{Prescribe})$ and a \emph{response} pattern $(\mathbf{G}(\text{Prescribe}\implies\mathbf{F}\text{Sign}))$ represented graphically in~\cite{ConDec:2006, Pesic2008} as shown in Fig.~\ref{fig:prescribesign}.
However, this approach suffers from the fact that the subsequent tools for execution and analysis will refer to the LTL expression (or further compilations to e.g. B\"uchi automata) and not the graphical notation. Also, the full generality of LTL may lead to a poor execution time.
 
This motivates researching the problem of finding an expressive declarative process language where both the constraints as well as the run time state can be easily visualized and understood by the end user and also 
allows an effective execution.
We believe that the declarative process model language of \emph{\modelname{}s} and its graphical representation 
proposed in this paper is a promising candidate. The model language is inspired
 by and a  conservative generalization of the declarative \emph{process matrix} model language~\cite{prohealth2008,DDBP:2008} used by our industrial partner and prime event structures~\cite{DBLP:conf/ac/Winskel86}. It is similar to~\cite{ConDec:2006, Pesic2008} in that it is based on a graph of constraints between events. The crucial difference is that only a fixed set of four primitive constraints is allowed and that the process semantics can be expressed directly as transitions between markings of the graph instead of via a translation to LTL.

We present distributed \modelname{}s as a sequence of three generalizations of prime event structures. A prime event structure  can be regarded as a minimal, declarative model for concurrent processes. It consists of a (possibly infinite) set of \emph{events} (that can happen at most once), a (partial order) \emph{causality relation} between events corresponding to the precedence LTL pattern above 
and a \emph{conflict relation} stating which events can not happen in the same execution. 

The first generalization, named \emph{condition response event structures},  is obtained by adding a \emph{response} relation between events and a set $\responses$ of \emph{initially required response events}. The initially required response events can be regarded as goals that must be fulfilled (or falsified) in order for an execution to be accepting. That is, for any event $e\in \responses$, either $e$ must eventually happen or it must become in conflict with an event that has happened in the past. The response relation in some sense corresponds to the response LTL pattern above as a dual relation to the usual causality relation: If an event $b$ is a response to an event $a$ then $b$ must happen at some point after event $a$ happens or become in conflict. However, note that the response pattern does not allow for conflicts. Operationally, as we will see in the following section, one can think of the event $b$ as being added to the set $\responses$ of required responses when $a$ happens.

Next we generalize condition response event structures by
 allowing each event to happen many times and replacing the symmetric conflict relation by an asymmetric relation which \emph{dynamically} determines which events are included in or excluded from the structure. To allow the graphs to represent intermediate run time state (e.g. like the marking of a Petri Net) 
 we also add sets $I$ and $E$ of respectively included and  executed events and refer to the triple of sets of pending responses, included and executed events as the \emph{marking} of the graph. This results in the model of \emph{Dynamic Condition Response Graphs}, short \modelshort.
 
 Finally, we reach the model of \emph{Distributed Dynamic Condition Response Graphs} allowing for role based \emph{distribution} by adding a set of \emph{principals} and
 a set of \emph{roles} assigned to both principals and events, and define that an event can only be executed  by a principal assigned one of the roles assigned to the event.
 
 Being based on  only four relations between events (condition, response, include, exclude) and the role assignment, the distributed dynamic condition response graphs can be simply visualized as a directed graph with a box for each event as nodes and four different kinds of arrows.  
  We base our graphical notation for the condition and response relations on the notation suggested in~\cite{DeclWFAalstPS09} for precendence and response LTL patterns, since they coincide when no events are excluded.  The inclusion and exclusion relations are denoted by arrows with a $+$ and $\%$ sign at the head respectively. We label each node with the activity of the event and add a small box to the top containing the roles that can execute the event. We annotate the graph by the marking, showing if an event is required as a response by adding a small exclamation mark, if it has happened in the past by a small check sign, and if it is excluded by making the box dashed. 
In addition we found it useful to show (by a small no-entry sign) if an event is blocked by an unfulfilled condition event, even though this information can be inferred from the condition relations and the currently included and executed events.
  We formalize the execution of \modelname{} as a labelled transition system, which is finite state if the graph is finite. Indeed, the states of transition system will be markings consisting of triples of sets of executed, included, and required response events. We define a (finite or infinite) run of the labelled transition system to be accepting if no response event is forever continuously included and pending without being executed. We end by characterizing the execution semantics by providing a mapping of \modelname{} to  B\"uchi-automata. 
 
 The rest of the paper is structured as follows. In Sec.~\ref{sect:cres} below we recall the definition of prime event structures and introduce condition response event structures as the first generalization. We show how the response relation allows to represent the notion of weak fairness. In Sec.~\ref{sect:dcrs} we introduce the model of \modelname{} (\modelshort{}) and distributed \modelshort{}. In Sec.~\ref{sect:buchi} we provide a mapping from \modelshort{} to B\"uchi-automata with $\tau$-events. In Sec.~\ref{sect:related} we briefly address related work. Finally, we conclude and discuss current and future work in Sec.~\ref{sect:conclusion}.
 
 This paper replaces and extends the work presented in the two previous short papers~\cite{places2010} and~\cite{tase2010}. The paper~\cite{places2010} introduced condition response event structures and dynamic condition response structures, which are essentially \modelname{} without markings.%
The paper~\cite{tase2010} provided a mapping from dynamic condition response structures to B\"uchi automata, but only capturing acceptance for the infinite runs. The mapping from \modelname s to 
 B\"uchi automata provided in the present paper characterizes also the acceptance of finite runs by introducing silent ($\tau$) transitions in the B\"uchi automata.

\section{Condition Response Event Structures}
\label{sect:cres}

As an intermediate step towards dynamic condition response graphs, we generalize prime event structures to allow for a notion of \emph{progress} based on a response relation. This model is interesting in itself as an extensional event-based model with progress, abstracting away from the intentional representation of repeated behavior. In particular we show that it allows for an elegant characterization of weakly fair runs of event structures. 

First let us recall the definition of a prime event structure and configurations of such~\cite{DBLP:conf/ac/Winskel86}.

\begin{definition}
A labeled \emph{prime event structure} (\es)  is a  5-tuple $E=(\events,\actions,\leq,\#,l)$
where 
\begin{enumerate}[(i)]
\item 
$\events$ is a (possibly infinite) set of events
\item $\actions$ is the set of actions
\item 
$\leq \ \subseteq \events\times \events$ is the \emph{causality} relation between events which is a partial order
\item 
$\# \  \subseteq  \events\times  \events$ is a binary \emph{conflict} relation between events which is irreflexive and symmetric
\item 
$l: \events\rightarrow \actions$ is the \emph{labeling} function mapping events to actions
\end{enumerate} 
The causality and conflict relations must satisfy the conditions that 
\begin{enumerate}
\item $\forall e,e',e''\in \events. e\#e'\leq e'' \implies e\# e''$ ,
\item $\forall e\in \events. e\downarrow =\{e'\mid e'< e\}$ is finite.
\end{enumerate} 
A \emph{configuration}  of $E$ is a set $c\subseteq \events$ of events satisfying the conditions

\begin{enumerate}
\item \emph{conflict-free}: $\forall e,e'\in c. \neg e\#e'$,
\item \emph{downwards-closed}: $\forall e\in c,e'\in \events. e'\leq e \implies e'\in c$.
\end{enumerate} 

A \emph{run} $\rho$ of $E$ is a (possibly infinite) sequence of labelled events $(e_0,l(e_0)),(e_1,l(e_1)),\ldots$ such that 
for all $i\geq 0. \cup_{0\leq j \leq i} \{e_j\}$ is a configuration. 

A run $(e_0,l(e_0)),(e_1,l(e_1)),\ldots$ is \emph{maximal} if any enabled event eventually happen or become in conflict, formally  $\forall e\in \events, i\geq 0. e\downarrow \subseteq (e_i\downarrow\cup\{e_i\})  \implies  \exists j\geq 0 .( e \# e_j \vee  e=e_j)$.  
\end{definition}
Action names $a \in \actions$ represent the actions the system might perform, an event  $e \in \events$ labelled with $a$  represents occurrence of action $a$ during the run of the system. The causality relation $e \leq e'$ means that event $e$ is a prerequisite for the event   $e'$ and the conflict relation $e  \#  e'$ implies that events $e$ and $e'$ both can not happen in the same run, more precisely one excludes the occurrence of the other.  
The definition of maximal runs follows the definition of weak fairness for concurrency models in~\cite{DBLP:conf/amast/Cheng95} and is equivalent to stating that the configuration defined by the events in the run is maximal with respect to inclusion of configurations.

We now generalize prime event structures to \emph{condition response event structures}, by
adding a dual  
 \emph{response} relation $\responserel$, such that $\{e'\mid e \responserel e'\}$  is the set of events that must happen (or be in conflict) after the event $e$ has happened for a run to be accepting. The resulting structures, named \emph{condition response event structures}, in this way add the possibility to state progress conditions.
 We also introduce a subset of the events $\responses$ of \emph{initial} responses, which are events that are initially required eventually to happen (or become in conflict). In this way the structures can represent the state after an event has been executed. As we will see below, it also allows us to capture the notion of maximal runs.

\begin{definition}
\textnormal{
A labeled \emph{\conreseventstructure} (\cres{}) over an alphabet $\actions$ is a  tuple $(\events,\responses,\actions, \conditionrel, \responserel,\#,l)$
where 
\begin{enumerate}[(i)]
\item  $(\events, \conditionrel,\#,l)$ is a labelled prime event structure, referred to as the underlying event structure
\item \textnormal{$\responserel \ \subseteq \events\times \events$ is the \emph{response} relation between events, satisfying that $\conditionrel\cup \responserel$ is acyclic}.
\item $\responses\subseteq \events$ is the set of \emph{initial responses}.
\end{enumerate} 
}
We define a configuration $c$ and run $\rho$ of a \cres{}  to be a respectively a configuration and run of the underlying event structure.
We define a run 
 $(e_0,l(e_0)),(e_1,l(e_1)),\ldots$  to be \emph{accepting} if $\forall e\in E, i\geq 0. e_i \responserel e \implies  \exists j\geq 0 .( e \# e_j \vee ( i<j \wedge e=e_j)$ and $\forall e\in R. \exists j\geq 0.( e \# e_j \vee e=e_j)$ . In words, any pending response event
 must eventually happen or be in conflict.  
 \end{definition}

A prime event structure can trivially be regarded as a \conreseventstructure{} with empty response relation. This provides an embedding of prime event structures into \conreseventstructure{}s which preserves configurations and runs.
 
 \begin{proposition}
The labelled prime event structure $(\events,\actions,\leq,\#,l)$ has the same runs as the \cres{} $(\events,\emptyset,\actions,\leq,\emptyset,\#,l)$ for which all runs are accepting. 
\end{proposition}

We can also embed event structures into CRES by considering every condition to be also a response and all events with no conditions to be initial responses. This characterizes the interpretation in~\cite{DBLP:conf/amast/Cheng95} where only \emph{maximal} runs are accepting. In other words, the embedding captures the notion of weakly fair execution of event structures.
\begin{proposition}
The labelled prime event structure $(\events,\actions,\leq,\#,l)$ has the same runs and \emph{maximal} runs as respectively the runs and the accepting runs of the \cres{} $(\events,\{e \mid e\downarrow =\emptyset\}, \actions,\leq,\leq,\#,l)$. 
\end{proposition}

\section{Distributed Dynamic Condition Response Graphs}
\label{sect:dcrs}

We now go on to generalize condition response event structures to dynamic condition response graphs (\dcrg). As opposed to event structures, a \modelname{} allows events to be executed multiple times
and there are no constraints on the condition and response relations. This allows for finite representations of infinite behavior, but also for introducing deadlocks. Moreover, the  conflict relation is generalized to two relations for dynamic exclusion and inclusion of events, which is more appropriate in a model where events can be re-executed and has shown useful in practice as a primitive for skipping events and constraints. 

\begin{definition}
A \emph{\modelname} is a tuple $G=(\events,\marking,\actions,\conditionrel, \responserel,\inex, l)$ 
where 
\begin{enumerate}[(i)]
\item 
$\events$ is the set of events
\item $\marking \in\markingset{}(G) = \power{\events}\times \power{\events}\times \power{\events}$ is the 
\emph{marking} and  $\markingset{}(G)$ is the set of all markings
\item $\actions$ is the set of actions
\item
$\conditionrel \subseteq \events \times $ E is the \emph{condition} relation
\item $\responserel \subseteq \events \times $ E is the \emph{response} relation
\item  $\inex : \events \times \events \rightharpoonup \{+,\%\}$ defines the \emph{dynamic inclusion/exclusion} relations by $ e \includerel  e'$ if $\inex(e, e') = + $  and  $ e \excluderel e'$ if $\inex(e, e') = \% $.  
\item $l:\events\rightarrow\actions$ is a labelling function mapping every event to an action.
\end{enumerate} 
We let \modelshort{} refer to the model of \modelname s.
\end{definition}

The  condition and response relations in \emph{\dcrg{}} are similar to the corresponding relations in \emph{\cres{}}, except that they are not constrained in any way. In particular, we may have cyclic relations.
The marking $\marking = (\executed, \responses, \included) \in \markingset{}(G)$ consists of three sets of events, capturing respectively which events have \emph{previously been executed} (\executed), which events are \emph{pending responses required to be executed or excluded} (\responses), and finally which events are currently \emph{included} (\included). The set of pending responses \responses{} of \modelshort{} thus plays the same role as the set of initial responses in the \cres{}.

 The \emph{dynamic inclusion/exclusion} relations $\includerel$ and $\excluderel$, represented by the (partial map) $\inex : \events \times \events \rightharpoonup \{+,\%\}$ , allow events to be included and excluded dynamically in the graph. The intuition is that only the currently included events are considered in evaluating the constraints. This means that if event $a$ has event $b$ as condition, but event $b$ is excluded from the graph then it is not required that $b$ has happened for $a$ to happen. Also, if event $a$ has event $b$ as response and event $b$ is excluded then it is not required that $b$ happens for the flow to be acceptable.
  Formally, the relation $ e \includerel e'$ expresses that, whenever event $e$ happens, it will include $e'$ in the graph. On the other hand,  $ e \excluderel e'$ expresses that when $e$ happens it  will exclude $e'$ from the graph.  
 
We  define the execution semantics of \dcrg{} by a labelled transition system with markings as states and define the set of accepting runs by requiring that no event must be continuously  included and pending.

\begin{definition}
\label{def:acceptingconditionfiniteruns}
For a \modelname{} $G=(\events,\marking,\actions,\conditionrel, \responserel,\inex,l)$ we define the corresponding labelled transition systems $T(G)$ to be the tuple 
$(\markingset{}(G),\marking{},
\rightarrow\subseteq \markingset{}(G)\times \actions \times \markingset{}(G))$ where $\markingset{}(G)$ 
is the set of markings of $G$,  $ \marking{} \in \markingset{}(G) $ is the initial marking, $ \rightarrow\subseteq \markingset{}(G)\times (\events\times\actions) \times {}(G) $ is the transition relation given by
 $\marking' \xrightarrow{(e,a)} \marking''   $
where 
\begin{enumerate}[(i)]
\item $\marking' = (\executed', \responses', \included')$ is the marking before transition
\item $\marking'' = (\executed'\cup \{e\}, \responses'', \included'')$ is the marking after transition
\item$e \in \included'$ and $l(e)=a$
\item$\{ e' \in \included' \mid  e' \conditionrel e\}  \subseteq \executed' $ \label{conditions}
\item$ \included''  =  ( \included'  \cup \{ e' \mid  e\includerel e'\}) \setminus \{ e' \mid   e \excluderel e' \} $
\item$ \responses'' = (\responses' \setminus \{ e \} ) \cup \{  e' \mid e \responserel e'  \} $ 
\end{enumerate}
We define a run $(e_0,a_0),(e_1,a_1),\ldots$ of the transition system 
to be a sequence of labels of a sequence of transitions $\marking_i \xrightarrow{(e_i,a_i)} \marking_{i+1} $, where $\marking_i = (\executed_i, \responses_i, \included_i)$ and $\marking_0 = \marking$.
We define a run to be accepting if $\forall i\geq 0, e\in \responses{}_i. \exists j \geq i. (e=e_j \vee e\not\in \included{}
_{j})$.
In words, 
a run is accepting 
if no response event is continuously included and pending without it happens. 
\end{definition}
The first two items in the above definition are markings before and after the transition. The third item expresses that only events $e$ that are currently included can be executed. The requirement saying that all currently included condition events for $e$ should have been executed previously is  expressed in~(\ref{conditions}).
The 
next two items are the updates to the sets of included events and pending responses respectively.    Note that an event $e'$ can not be both included and excluded by the same event $e$, but an event may exclude itself. Also an event may trigger itself as a response and/or has itself as condition.

If one only want to consider finite runs, which is common for workflows, the acceptance condition degenerates to requiring that no pending response is included at the end of the run. This corresponds to defining all states where $\responses\cap \included=\emptyset$ to be accepting states and define the accepting runs to be those ending in an accepting state.
 If infinite runs are also of interest (as e.g. for reactive systems and the LTL logic) the acceptance condition can be captured by a mapping to a B\"uchi-automaton with $\tau$-events which we give
  in Sec.~\ref{sect:buchi} below.

\begin{figure}[htp]
 \vspace{-0.5cm}
\centering
\subfigure[\# relation in \cres]{\label{fig:conflictincres}\includegraphics[width=0.4\textwidth]{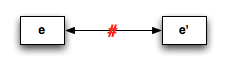}}
\hspace{0.75in}
\subfigure[Encoding of \# in \dcrg]{\label{fig:conflictindcr}\includegraphics[width=0.4\textwidth]{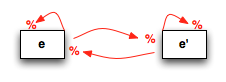}}    
\vspace{-.4cm}\caption{Encoding conflicting events in \cres{} as mutual excluding events in \dcrg }
\label{fig:cerstodcrsexample}
\end{figure}

 A \cres{} can be represented as a \modelname{} by making every event exclude itself and encode the 
 conflict relation by defining any two conflicting events to mutually exclude each other as shown in figure~\ref{fig:conflictindcr}. 
  
\begin{proposition}
The \cres{} $(\events,\responses, \actions,\conditionrel,\responserel,\#,l)$ has the same runs and accepting runs as the \modelname{} 
$(\events, \marking,\actions,\conditionrel,\responserel,\inex,l)$
where $\marking{} = (\emptyset,\responses, \events), \inex(e,e')=\%$ if $e=e'$ or $e\# e'$ and undefined otherwise.
\end{proposition}

We now define  \emph{distributed} \modelname{}s by adding roles and principals.
\begin{definition}
A \emph{distributed} \modelname{} 
is a tuple (G, \roles, \agents, \as) 
where 
\begin{enumerate}
\item $G = (\events,\marking, \actions,\conditionrel,  \responserel,\inex,l)$ is a \modelname{},
 \item  \roles{} is a set of \emph{roles}, 
 \item \agents{} is a set of \emph{principals} (e.g. persons or processors) and 
 \item $\as\subseteq (\agents \cup \actions) \times \roles$ is the role assignment relation to principals and actions. 
 \end{enumerate}
\end{definition}

 For a \emph{distributed} \dcrg{}, the role assignment relation indicates the roles (access rights) assigned to principals  and which roles gives right to execute which actions. As an example, assume $Peter\in \agents$ and $Doctor\in \roles$, then if $Peter \as Doctor$ and $Sign \as Doctor$ then $Peter$ as a doctor can sign as a doctor. 
 
This is formalized by defining the labelled transition semantics for a  \emph{distributed} \modelname{} 
$D = (G, \roles, \agents, \as) $
to have the same states as the underlying \modelname{} $G$, and the transitions $ \rightarrow\subseteq \markingset{}(G) \times \events\times(\agents\times\actions\times\roles) \times \markingset{}(G)$ defined by $\marking' \xrightarrow{(e,(p,a,r))} \marking''$ if $p \as r$ and $a \as r$ and $\marking' \xrightarrow{(e,a)} \marking''$ in the underlying \modelname. We define a run to be (finite or infinite) sequence of labels $(e_0,(p_0,a_0,r_0))(e_1,(p_1,a_1,r_1))\ldots$ of a sequence of transitions 
$\marking_{i} \xrightarrow{(e_i,(p_i,a_i,r_i))} \marking_{i+1}$
starting from the initial marking. We define a run to be accepting if the underlying run of the \dcrg{} is accepting.
 
\begin{figure}[htp]
\vspace{-.5cm}
\centering
\subfigure[Prescribe Medicine Example]{\label{fig:ordinatemedicine1}\includegraphics[width=0.33\textwidth]{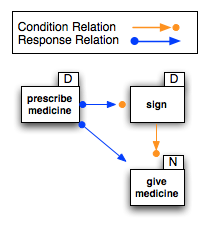}}
\hspace{0.75in}
\subfigure[Prescribe Medicine Example With Check]{\label{fig:ordinatemedicine2}\includegraphics[width=0.42\textwidth]{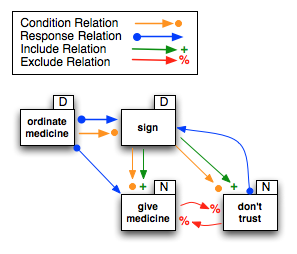}}    
\vspace{-.5cm}
\caption{\modelshort{} example in graphical notation }
\label{fig:dcrsexample}
\vspace{-.5cm}
\end{figure}

We are now ready to give the small example workflow from the introduction graphically as a distributed \modelname{} shown in Fig.~\ref{fig:ordinatemedicine1}.
It contains three events: \textsf{prescribe medicine} (the doctor calculates and writes the dose for the medicine), \textsf{sign} (the doctor certifies the correctness of the calculations) and \textsf{give medicine} (the nurse administers medicine to patient).  The events are also labelled by the assigned roles (\textsf{D} for Doctor and \textsf{N} for Nurse).

The arrow \condresponserel{}  between \textsf{prescribe medicine} and \textsf{sign} indicates that the two events are related by both the condition relation and the response relation. The condition relation means that the \textsf{prescribe medicine} event must happen at least once before the \textsf{sign} event. The %
response relation 
enforces that, if the \textsf{prescribe medicine} event happen, subsequently at some point the \textsf{sign} event must happen for the run to be accepted. Similarly, the response relation between \textsf{prescribe medicine} and \textsf{give medicine} 
enforces that, if the \textsf{prescribe medicine} event happen, subsequently at some point the \textsf{give medicine} event must happen for the flow to be accepted.

\begin{wrapfigure}{R}{0.40\textwidth}
  \vspace{-20pt}
  \begin{center}
    \includegraphics[width=0.40\textwidth]{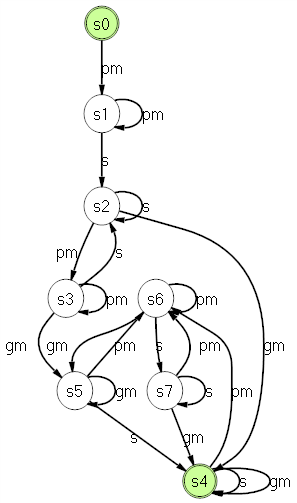}
  \end{center}
  \vspace{-20pt}
  \caption{Transition system for DCR graph from Fig~\ref{fig:ordinatemedicine1}}
  \label{fig:givemedicine-finite-statespace}  
  \vspace{-10pt}
\end{wrapfigure}

Finally, the condition relation between \textsf{sign} and \textsf{give medicine} enforces that the signature event must have happened before the medicine can be given.  Note the nurse can give medicine many times, and that the doctor can at any point choose to prescribe new medicine and sign again. (This will not block the nurse from continue to give medicine. The interpretation is that the nurse may have to keep giving medicine according to the previous prescription). The transition system for 
 the prescribe medicine
  example 
   is shown in Fig.~\ref{fig:givemedicine-finite-statespace}. For simplicity we only show the actions as labels. The green states are the states with no included pending responses.  

The dynamic inclusion and exclusion of events is illustrated by an extension to the scenario (also taken from the real case study):
If the nurse distrusts the prescription by the doctor, it should be possible to indicate it, and this action should force either a new prescription followed by a new signature or just a new signature. As long the new signature has not been added, medicine must not be given to the patient.

This scenario can be modeled as shown in  Fig.~\ref{fig:ordinatemedicine2}, where one more event labelled  \textsf{don't trust} is added. Now, the nurse have a choice to indicate distrust of prescription and thereby exclude \textsf{give medicine} until the doctor execute \textsf{sign} again. Executing \textsf{don't trust} action will 
 make \textsf{sign}  a pending response.  So the only way to reach an accepting run  is to re-execute \textsf{sign} which will include \textsf{give medicine}. The doctor may choose to re-do \textsf{prescribe medicine} followed by \textsf{sign} (if the reason for distrusting the prescription was indeed valid) or simply re-do \textsf{sign}.

\begin{figure}[htp]
\vspace{-.5cm}
\centering
\subfigure[Prescribe Medicine Example]{\label{fig:dcrsruntime1}\includegraphics[width=0.49\textwidth]{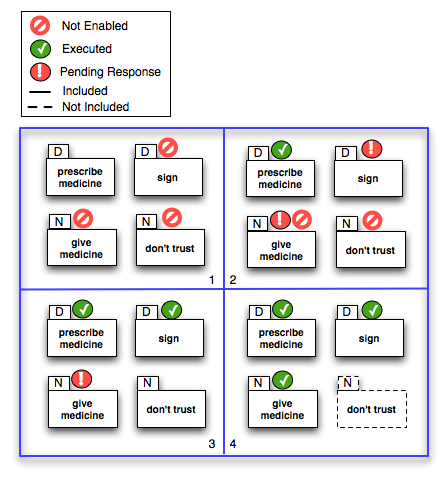}}
\hspace{0.0in}
\subfigure[Prescribe Medicine Example \textsf{Don't trust}]{\label{fig:dcrsruntime2}\includegraphics[width=0.495\textwidth]{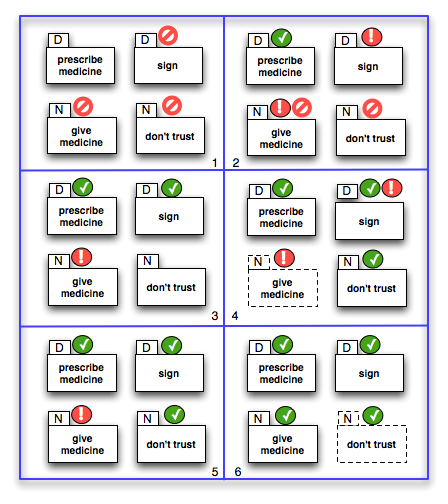}}    
\vspace{-.5cm}\caption{\modelshort{} Runtime state graphical notation }
\label{fig:dcrsexampleruntime}
\vspace{-.5cm}
\end{figure}

In Fig.~\ref{fig:dcrsexampleruntime} below we propose a graphical notation that illustrates the run-time information during two different runs of the extended scenario in Fig.~\ref{fig:ordinatemedicine2}. We 
 use three different small icons (\O, $\surd$,!) above the boxes to show if the event is not enabled (i.e. it is blocked by an included condition event that has not been executed), if it has been executed (i.e. included in the set $E$ in the marking), and if it is required as a response (i.e. included in the set $R$ in the marking). We indicate that an event is excluded (i.e. not included in the set $I$ in the marking) by making the box around the event dashed. 

Fig~\ref{fig:dcrsruntime1} shows four states of a run in the workflow process in Fig.~\ref{fig:ordinatemedicine2}, starting in the initial state where all events except \textsf{prescribe medicine} is blocked. The second state is the result of executing \textsf{prescribe medicine}, now showing that \textsf{sign} and \textsf{give medicine} are required as responses and that \textsf{sign} is no longer blocked. The third state is the result of executing the \textsf{sign} event, which enables \textsf{give medicine} and \textsf{don't trust}. Finally, the fourth state is the result of executing the \textsf{give medicine} event, excluding the \textsf{don't trust} event.

Similarly, Fig.~\ref{fig:dcrsruntime2} shows the six states of a run where the nurse executes \textsf{don't trust} in the third step, leading to a different fourth state where \textsf{give medicine} is excluded (but still required as response if it gets included again) and \textsf{sign} is required as response. The fifth state shows the result of the doctor executing \textsf{sign}, which re-includes \textsf{give medicine}, which is then executed, leading to the final state where all events have been executed, and \textsf{don't trust} is excluded.

\section{From \modelshort{} to B\"uchi-automata}
In this section, we show how to characterize the acceptance condition for \modelshort{} by a mapping to 
 B\"uchi-automata with $\tau$-event. Recall that a B\"uchi-automaton is a finite state automaton accepting only infinite runs, and only the runs that pass through an accepting state infinitely often. Acceptance of finite runs can be represented in the standard way by introducing a special silent event, e.g. a $\tau$-event, which may be viewed as a delay. If an infinite accepting run contains infinitely many delays it then represent an accepting run containing only a finite number of (real) events. We define a B\"uchi-automaton with $\tau$-event as follows.

\begin{definition}
A \emph{B\"uchi-automaton with $\tau$-event} is a tuple $(S,s, Ev_{\tau},\rightarrow\subseteq S\times Ev_{\tau} \times S, F)$ where $S$ is the set of \emph{states}, $s \in S $ is the \emph{initial state}, $Ev_{\tau}$ is the set of \emph{events} containing the special event $\tau$, $ \rightarrow\subseteq S\times Ev_{\tau} \times S$ is the \emph{transition relation}, and $F$ is the set of \emph{accepting states}. A (finite or infinite) run is a sequence of labels not containing the $\tau$ event that can be obtained by removing all $\tau$ events from a sequence of labels of transitions starting from the initial state. The run is \emph{accepting}  if the sequence of transitions 
passes through an accepting state infinitely often.
\end{definition}

 The mapping from \modelshort{} to B\"uchi-automata is not entirely trivial, since we at any given time may have several pending responses and thus must make sure that all of them are eventually executed or excluded. To make sure we progress, we assume any fixed order of the finite set of events $E$ of the given \modelname{} and enforce the execution (or exclusion) of response events in that order. For an event $e\in E$ we write $rank(e)$ for its rank in that order and for a subset of events $E'\subseteq E$ we write $min(E')$ for the event in $E'$ with the minimal rank.

 \begin{definition}
\textnormal{
For $G=(\events,\marking,\actions,\conditionrel, 
\responserel,\inex,l,\roles,\agents, \as)$ a finite distributed \modelname{}  where $\events=\{e_1,\ldots,e_n\}$ and $rank(e_i)=i$, we define the corresponding 
 B\"uchi-automaton with $\tau$-event to be the tuple  $B(G)=(S,s,\rightarrow\subseteq S\times Ev_{\tau} \times S, F
)$ where \begin{itemize}
\item $S= \markingset{}(G)\times \{1,\ldots,n\}\times \{0,1\}$ is the set of states, 
\item $Ev_{\tau}=(\events\times(\agents\times\actions\times\roles))\cup\{\tau\}$ is the set of events, 
\item $s=(\marking,1,1)$ if $I \cap \responses =\emptyset$, and $s=(\marking,1,0)$ otherwise
\item $F=\markingset{}(G)\times \{1,\ldots,n\}\times \{1\}$ is the set of accepting states 
and
\item$ \rightarrow\subseteq S\times Ev_{\tau} \times S$ is the transition relation given by
\end{itemize}
 \center
  {$(\marking',i,j) \xrightarrow{\mbox{\ \  $\tau$  \ \ }} (\marking', i, j')   $}
  where
\begin{enumerate}[(a)]
\item $j'=1$ if \label{acceptance_a} $\included'\cap \responses'=\emptyset$  \label{noincludedresponse}
otherwise $j'=0$.
 \end{enumerate}
 \center{and}
 \center
  {$(\marking',i,j) \xrightarrow{\mbox{\ \  (e,(p,a,r))  \ \ }} (\marking'',i',j')   $}
where 
\begin{enumerate}[(i)]
\item $\marking' = (\executed', \responses', \included')$ and $\marking'' = (\executed' \cup \{e\}, \responses'', \included'')$ 
\item $\marking' \xrightarrow{\mbox{\ \  (e,(p,a,r))  \ \ }} \marking''  $ is a transition of $T(D)$
\item $j'=1$ if \label{acceptance}
\begin{enumerate}
\item $\included'' \cap \responses''=\emptyset$ or \label{noincludedresponse}
\item $min(M_{r})\in  (\included'\cap \responses' \backslash (\included''\cap \responses''))\cup\{e\}$ or \label{minaboveexecutedorexcluded}
\item $M_{r}=\emptyset$ and $min(\included'\cap \responses')\in (\included'\cap \responses' \backslash (\included''\cap \responses''))\cup\{e\}$ \label{minexecutedorexcluded}
\end{enumerate}
otherwise $j'=0$.
\item $i'=rank(min(M_{r}))$ if $min(M_{r})\in  (\included'\cap \responses' \backslash (\included''\cap \responses''))\cup\{e\}$ or else \label{newrankfirst}
\item $i'=rank(min(\included' \cap \responses'))$ if $M_{r}=\emptyset$ and $min(\included'\cap \responses')\in (\included'\cap \responses' \backslash (\included''\cap \responses''))\cup\{e\}$ or else \label{newranksec}
\item $i'=i$ otherwise. \label{nonewrank}
\end{enumerate}
for $M_{r}=\{e\in \included'\cap \responses' \mid rank(e)\mathrel{>} i\}$.
}
\end{definition}
In the marking $\marking$, the set $\executed$ records the events that have been executed, where as $\included$ and $\responses$ records the events that are currently included and pending responses respectively.
The index $i$ is used  to make sure that no event stays forever included and in the pending response set without being executed. Finally, the flag $j$ indicates if the state is accepting or not.
The conditions (\ref{acceptance_a}) and (\ref{acceptance}) define when a state is accepting. 
Either there are no included pending responses in the resulting state (\ref{noincludedresponse}) or  the included pending response with the minimal rank above the index $i$ was either excluded or executed (\ref{minaboveexecutedorexcluded}). Alternatively, if the set of included pending responses with rank above the index $i$ is empty and the included pending response with the minimal rank is excluded or executed (\ref{minexecutedorexcluded}), then also the resulting state will be accepting. 
Condition (\ref{newrankfirst}) records the new rank if the resulting state is accepting according to condition (\ref{minaboveexecutedorexcluded}) and similarly when the state is accepting according to condition (\ref{minexecutedorexcluded}), the condition (\ref{newranksec}) records the new rank. 

\begin{figure}[h!]
\vspace{-.2cm}
\centering
\includegraphics[width=0.75\textwidth]{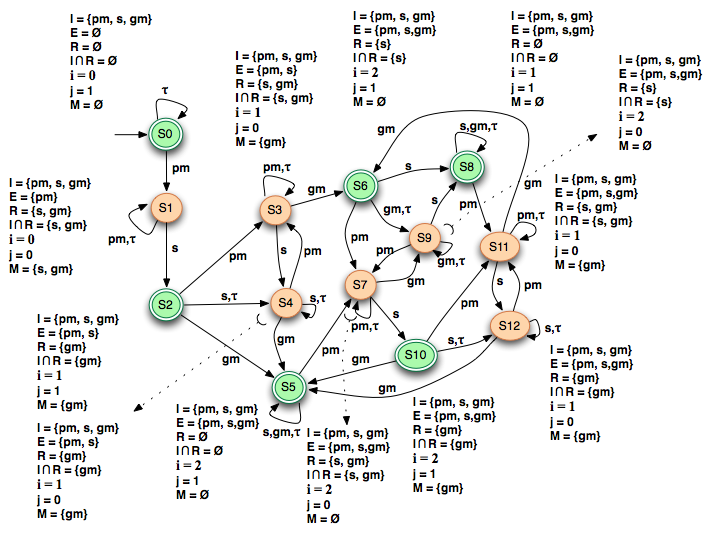}
\vspace{-.5cm}
\caption{The B\"uchi-automaton for DCR Graph from Fig.~\ref{fig:ordinatemedicine1} annotated with state information }
\label{fig:givemedicinebuchi}
\vspace{-.3cm}
\end{figure}

\begin{wrapfigure}{R}{0.50\textwidth}
  \vspace{-20pt}
  \begin{center}
    \includegraphics[width=0.50\textwidth]{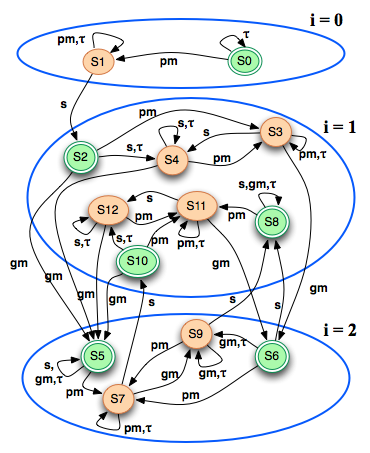}
  \end{center}
  \vspace{-20pt}
  \caption{The B\"uchi-automaton with stratified view}
  \label{fig:givemedicinebuchimultiplecopies}  
  \vspace{-10pt}
\end{wrapfigure}

To give a simple example of the mapping, let us consider the \modelname{} in Fig.~\ref{fig:ordinatemedicine1} and the corresponding B\"uchi-automaton in Fig.~\ref{fig:givemedicinebuchi}.

The key point to note is that the automaton enters an accepting state if there is no pending responses, or if the pending response which is the minimal ranked event according to the index $i$ is executed or excluded. State $S7$ and $S11$ illustrate the use of the rank: Both states have the two events \textsf{s} (having rank 1) and \textsf{gm} as pending responses. In state $S7$ only executing event \textsf{s} leads to an accepting state ($S10$). The result of executing event \textsf{gm} is to move to state $S9$ which is not accepting. Dually, in state  $S11$ only executing event \textsf{gm} leads to an accepting state ($S16$). The result of executing event \textsf{s} is to move to state $S12$ which is not accepting.
Fig.~\ref{fig:givemedicinebuchimultiplecopies} shows a stratified view of the automaton, dividing the state sets according to the rank $i$ in order to emphasize the role of the rank in guaranteeing progress.

We end by stating the main theorem that the mapping from \modelname{} to B\"uchi-automata characterizes the execution semantics.

\begin{theorem}
For a finite distributed \modelname{} $D$ the B\"uchi-automaton with $\tau$-event $B(D)$ has the same runs and accepting runs as $D$.
\end{theorem}

\label{sect:buchi}

\section{Related Work}
\label{sect:related}
There exists many different approaches to formally specify and enact business  processes and workflows. As it is not possible to provide a complete overview of all related work, we give here just a brief overview of some of the formalisms which are related to our work and compare them to \modelshort.

As already described in the introduction, the authors in~\cite{ConDec:2006,DeclWFAalstPS09} have proposed $ConDec$,  a declarative language for modeling and enacting the dynamic business processes based on Linear Temporal Logic (LTL). In~\cite{DecSerFlow:2006}, the authors have proposed Declarative Service Flow language  (DecSerFlow) to specify, enact and monitor service flows, which is a sister language for $ConDec$. Both the languages share the same concepts and are supported in the $Declare$~\cite{DeclWFAalstPS09} tool. They specifies $what$ should be done, instead of specifying $how$ it should be done, there by leaving more flexibility to users. 
The enactment in both the languages is defined by translating the constraints specified in LTL, into a Buchi automaton and executing the workflow/service by executing the referring Buchi automaton.  LTL being a very expressive language, the Declare tool suffers from efficiency problems in executing models with large specification~\cite{DeclWFAalstPS09}. Even though our approach is related to the  work in~\cite{ConDec:2006,DecSerFlow:2006}, \modelshort{} has a fixed set of constraints that makes it simpler to learn and possible to describe the execution semantics directly as transitions between markings of the graph.

The $Event\ Calculus$~\cite{eventcalculusworkflows,Cicekli20062227,eventcalculusKR} is another logic-based methodology for specification and execution of workflows. It is a logic programming formalism for representing events and their effects in the context of database applications. The authors have expressed the basic control flow primitives as a set of logical formulas and used  axioms of $Event\ Calculus$ to specify activity dependency execution and agent assignments rules. Their workflow model also supports enactment 
and iteration of activities, but does not support verification of global and temporal constraints on workflow activities. Also, their approach is limited to imperative/procedural workflow  modeling languages.

Concurrent Transaction Logic (CTR) is used in ~\cite{Davulcu98logicbased} as a language for specifying, analysis, scheduling and verification of workflows. 
The authors have used CTR formulas for expressing the local and global properties of workflows. Reasoning about the workflows has been done with the help of proof theory and semantics of logic. 
In ~\cite{Senkul02alogical}, the authors have used Concurrent Constraint Transaction Logic (CCTR) which is a flavor of CTR integrated with Constraint Logic Programming for scheduling workflows. 
Like the other logic programming systems, the authors in  ~\cite{Davulcu98logicbased,Senkul02alogical} have used the proof theory of CTR as run-time environment for enactment of workflows.  The CTR approach mainly aims at developing an algorithm for consistency checking and verification of properties of workflows, but only limited to imperative modeling languages. 

Petri nets have 
been studied and used extensively in the domain of workflows and business processes, see e.g.~\cite{vanderAalst:1998:APN,Desel2000, Aalst2011}. The correspondence between event structures and Petri nets are well studied, see e.g.~\cite{handbookmodelsforconcurrency}. It is however not possible to express the notion of pending responses directly in standard Petri Nets.

Our formal model \dcrg{} also relates to the declarative approaches used by \emph{Guard-Stage-Milestone} (GSM) model~\cite{Hull2010}  by Hull et al, presented as an invited talk at the WS-FM 2010 workshop. 
The GSM meta-model uses declarative approach for specification of life cycles, which is part of research on data driven artifact-centric business processes~\cite{Bhattacharya07towardsformal,datacentricbusinessprocesses:2009,businessartifacts:2009}, carried out by the IBM Research. The operational semantics of GSM  are based on Event-Condition-Action (ECA) rules, and provide a basis for formal verification and reasoning.

\section{Conclusion and Future Work}
\label{sect:conclusion}
We have presented \modelname s, short \modelshort, as a new declarative, event-based workflow process model inspired by the workflow language employed by our industrial partner~\cite{DDBP:2008}. We have demonstrated the use and flexibility of the model on a small example taken from a field study on danish hospitals~\cite{prohealth2008} and proposed a graphical notation for presenting both the processes and their run-time state.

The model was presented as a sequence of generalizations of the classical model for concurrency of prime event structures~\cite{DBLP:conf/ac/Winskel86}. The first generalization introduced a notion of progress to event structures by replacing the usual causal order by two dual relations, a \emph{condition} relation $\conditionrel$ expressing for each event which events it has as preconditions  and a \emph{response} relation $\responserel$ expressing for each event which events that must happen (or become in conflict) after it has happened. We demonstrated that the resulting model, named \emph{condition response event structures}, 
 can express the standard notion of weak concurrency fairness.

The next generalization is to allow for finite representations of infinite behaviours by allowing
\emph{multiple execution}, and \emph{dynamic inclusion} and \emph{exclusion} of events, resulting in the model of \emph{\modelname s}. 
 
Finally,  we extended the model to allow distribution of events via roles and presented a graphical notation inspired by related work by van der Aalst et al.~\cite{ConDec:2006,DeclWFAalstPS09}, but extended to include information about the run-time state (e.g. markings).

We prove that all generalizations conservatively contain the previous model. Moreover, we provide a mapping from \modelname s
 to B\"uchi-automata characterising the acceptance condition for finite and infinite runs.

One key advantage of the \modelname{}s  compared to the related work  explored in
~\cite{ConDec:2006,DeclWFAalstPS09, Davulcu98logicbased, Cicekli20062227} is that the latter logics are 
more complex to visualize and understand by people not trained in logic.  Another advantage, illustrated in the given mapping to B\"uchi-automata and our graphical visualization of the run time state, is that the execution of \modelname{}s can be based on a relatively simple information about the run-time state, which can also be visualized directly as annotations (marking) on the graph. We have implemented a prototype engine and mapping to the input format for the SPIN model checker, and are currently working on implementing a simulator for DCRS able to visualize the state graphically in this way.

Current and future work include studying  extensions of the \modelshort{} model with time, exceptions, nesting (sub processes) and data (as publish/subscribe events to changes of data) and the relationship between \modelshort{} and Petri Net with time and infinite behavior~\cite{PetriFair}. Also we are investigating the expressiveness of the model compared to LTL and the work in~\cite{ConDec:2006}. Along the line of work in~\cite{handbookmodelsforconcurrency} we investigate the definition of interfaces between concurrently interacting \modelname s, a (categorical) theory of simulations and defining unfolding (forgetful) mappings from \modelshort{} to \cres{} and from \cres{} to event structures.
We expect the theory to be useful in achieving compositional design and verification of workflow processes, as well as studying the impact of adapting or adding new interacting workflow processes to a pool of processes. Finally we intend to explore the relation to the recent work on event-based business processes.
\subsection*{Acknowlegments}
This research is supported by the Trustworthy Pervasive Healthcare Services (TrustCare) project and the Computer Supported Mobile Adaptive Business Processes (CosmoBiz) project. Danish Research Agency, Grants $\#$ 2106-07-0019 and $\#$ 274-06-0415 (www.TrustCare.eu and www.CosmoBiz.dk).

%
\label{sect:bib}
\bibliographystyle{eptcs} 
\bibliography{bibnewEPTCS}

\begin{thebibliography}{10}
\providecommand{\bibitemdeclare}[2]{}
\providecommand{\urlprefix}{Available at }
\providecommand{\url}[1]{\texttt{#1}}
\providecommand{\href}[2]{\texttt{#2}}
\providecommand{\urlalt}[2]{\href{#1}{#2}}
\providecommand{\doi}[1]{doi:\urlalt{http://dx.doi.org/#1}{#1}}
\providecommand{\bibinfo}[2]{#2}

\bibitemdeclare{article}{Aalst2011}
\bibitem{Aalst2011}
\bibinfo{author}{W.~van~der Aalst}, \bibinfo{author}{K.~van Hee},
  \bibinfo{author}{A.~ter Hofstede}, \bibinfo{author}{N.~Sidorova},
  \bibinfo{author}{H.~Verbeek}, \bibinfo{author}{M.~Voorhoeve} \&
  \bibinfo{author}{M.~Wynn} (\bibinfo{year}{2011}):
  \emph{\bibinfo{title}{Soundness of workflow nets: classification,
  decidability, and analysis}}.
\newblock {\sl \bibinfo{journal}{Formal Aspects of Computing}}
  \bibinfo{volume}{23}, pp. \bibinfo{pages}{333--363},
  \doi{10.1007/s00165-010-0161-4}.

\bibitemdeclare{article}{vanderAalst:1998:APN}
\bibitem{vanderAalst:1998:APN}
\bibinfo{author}{W.~M.~P. van~der Aalst} (\bibinfo{year}{1998}):
  \emph{\bibinfo{title}{The Application of {Petri} Nets to Workflow
  Management}}.
\newblock {\sl \bibinfo{journal}{The Journal of Circuits, Systems and
  Computers}} \bibinfo{volume}{8}(\bibinfo{number}{1}), pp.
  \bibinfo{pages}{21--66}, \doi{10.1142/S0218126698000043}.

\bibitemdeclare{article}{DeclWFAalstPS09}
\bibitem{DeclWFAalstPS09}
\bibinfo{author}{Wil M.~P. van~der Aalst}, \bibinfo{author}{Maja Pesic} \&
  \bibinfo{author}{Helen Schonenberg} (\bibinfo{year}{2009}):
  \emph{\bibinfo{title}{Declarative workflows: Balancing between flexibility
  and support}}.
\newblock {\sl \bibinfo{journal}{Computer Science - R{\&}D}}
  \bibinfo{volume}{23}(\bibinfo{number}{2}), pp. \bibinfo{pages}{99--113},
  \doi{10.1007/s00450-009-0057-9}.

\bibitemdeclare{inproceedings}{ConDec:2006}
\bibitem{ConDec:2006}
\bibinfo{author}{Wil~M.P van~der Aalst} \& \bibinfo{author}{Maja Pesic}
  (\bibinfo{year}{2006}): \emph{\bibinfo{title}{A Declarative Approach for
  Flexible Business Processes Management}}.
\newblock In: {\sl \bibinfo{booktitle}{Proceedings of DPM 2006}},
  \bibinfo{series}{LNCS}, \bibinfo{publisher}{Springer Verlag}.

\bibitemdeclare{inproceedings}{DecSerFlow:2006}
\bibitem{DecSerFlow:2006}
\bibinfo{author}{Wil~M.P van~der Aalst} \& \bibinfo{author}{Maja Pesic}
  (\bibinfo{year}{2006}): \emph{\bibinfo{title}{{D}ec{S}er{F}low: Towards a
  Truly Declarative Service Flow Language}}.
\newblock In \bibinfo{editor}{M.~Bravetti}, \bibinfo{editor}{M.~Nunez} \&
  \bibinfo{editor}{Gianluigi Zavattaro}, editors: {\sl
  \bibinfo{booktitle}{Proceedings of Web Services and Formal Methods (WS-FM
  2006)}}, {\sl \bibinfo{series}{LNCS}} \bibinfo{volume}{4184},
  \bibinfo{publisher}{Springer Verlag}, pp. \bibinfo{pages}{1--23}.

\bibitemdeclare{inproceedings}{Bhattacharya07towardsformal}
\bibitem{Bhattacharya07towardsformal}
\bibinfo{author}{Kamal Bhattacharya}, \bibinfo{author}{Cagdas Gerede},
  \bibinfo{author}{Richard Hull}, \bibinfo{author}{Rong Liu} \&
  \bibinfo{author}{Jianwen Su} (\bibinfo{year}{2007}):
  \emph{\bibinfo{title}{Towards Formal Analysis of Artifact-Centric Business
  Process Models}}.
\newblock In \bibinfo{editor}{Gustavo Alonso}, \bibinfo{editor}{Peter Dadam} \&
  \bibinfo{editor}{Michael Rosemann}, editors: {\sl
  \bibinfo{booktitle}{Business Process Management}}, {\sl
  \bibinfo{series}{Lecture Notes in Computer Science}} \bibinfo{volume}{4714},
  \bibinfo{publisher}{Springer Berlin / Heidelberg}, pp.
  \bibinfo{pages}{288--304}, \doi{10.1007/978-3-540-75183-0}.

\bibitemdeclare{inproceedings}{agentcoorddbsystems}
\bibitem{agentcoorddbsystems}
\bibinfo{author}{Christoph Bussler} \& \bibinfo{author}{Stefan Jablonski}
  (\bibinfo{year}{1994}): \emph{\bibinfo{title}{Implementing agent coordination
  for workflow management systems using active database systems}}.
\newblock In: {\sl \bibinfo{booktitle}{Research Issues in Data Engineering,
  1994. Active Database Systems. Proceedings Fourth International Workshop
  on}}, pp. \bibinfo{pages}{53--59}, \doi{10.1109/RIDE.1994.282853}.

\bibitemdeclare{inproceedings}{DBLP:conf/amast/Cheng95}
\bibitem{DBLP:conf/amast/Cheng95}
\bibinfo{author}{Allan Cheng} (\bibinfo{year}{1995}):
  \emph{\bibinfo{title}{Petri Nets, Traces, and Local Model Checking}}.
\newblock In: {\sl \bibinfo{booktitle}{Proceedings of AMAST}}, pp.
  \bibinfo{pages}{322--337}.

\bibitemdeclare{article}{Cicekli20062227}
\bibitem{Cicekli20062227}
\bibinfo{author}{Nihan~Kesim Cicekli} \& \bibinfo{author}{Ilyas Cicekli}
  (\bibinfo{year}{2006}): \emph{\bibinfo{title}{Formalizing the specification
  and execution of workflows using the event calculus}}.
\newblock {\sl \bibinfo{journal}{Information Sciences}}
  \bibinfo{volume}{176}(\bibinfo{number}{15}), pp. \bibinfo{pages}{2227 --
  2267}, \doi{10.1016/j.ins.2005.10.007}.
\newblock
  \urlprefix\url{http://www.sciencedirect.com/science/article/B6V0C-4HMGKHK-1/%
2/8d83da5d71878f08a894fa8b9deb9933}.

\bibitemdeclare{incollection}{eventcalculusworkflows}
\bibitem{eventcalculusworkflows}
\bibinfo{author}{Nihan~Kesim Cicekli} \& \bibinfo{author}{Yakup Yildirim}
  (\bibinfo{year}{2000}): \emph{\bibinfo{title}{Formalizing Workflows Using the
  Event Calculus}}.
\newblock In: {\sl \bibinfo{booktitle}{Proceedings of the 11th International
  Conference on Database and Expert Systems Applications}},
  \bibinfo{series}{DEXA '00}, \bibinfo{publisher}{Springer-Verlag},
  \bibinfo{address}{London, UK}, pp. \bibinfo{pages}{222--231}.
\newblock \urlprefix\url{http://portal.acm.org/citation.cfm?id=648313.761618}.

\bibitemdeclare{article}{businessartifacts:2009}
\bibitem{businessartifacts:2009}
\bibinfo{author}{David Cohn} \& \bibinfo{author}{Richard Hull}
  (\bibinfo{year}{2009}): \emph{\bibinfo{title}{Business Artifacts: A
  Data-centric Approach to Modeling Business Operations and Processes}}.
\newblock {\sl \bibinfo{journal}{IEEE Data Eng. Bull.}}
  \bibinfo{volume}{32}(\bibinfo{number}{3}), pp. \bibinfo{pages}{3--9}.
\newblock \urlprefix\url{http://sites.computer.org/debull/A09sept/david.pdf}.

\bibitemdeclare{inproceedings}{Davulcu98logicbased}
\bibitem{Davulcu98logicbased}
\bibinfo{author}{Hasam Davulcu}, \bibinfo{author}{Michael Kifer},
  \bibinfo{author}{C.~R. Ramakrishnan} \& \bibinfo{author}{I.V. Ramakrishnan}
  (\bibinfo{year}{1998}): \emph{\bibinfo{title}{Logic Based Modeling and
  Analysis of Workflows}}.
\newblock In: {\sl \bibinfo{booktitle}{Proceedings of ACM
  SIGACT-SIGMOD-SIGART}}, \bibinfo{publisher}{ACM Press}, pp.
  \bibinfo{pages}{1--3}.

\bibitemdeclare{incollection}{Desel2000}
\bibitem{Desel2000}
\bibinfo{author}{J~Desel} \& \bibinfo{author}{Thomas Erwin}
  (\bibinfo{year}{2000}): \emph{\bibinfo{title}{Modeling, Simulation and
  Analysis of Business Processes}}.
\newblock In \bibinfo{editor}{Wil van~der Aalst}, \bibinfo{editor}{J~Desel} \&
  \bibinfo{editor}{Andreas Oberweis}, editors: {\sl
  \bibinfo{booktitle}{Business Process Management}}, {\sl
  \bibinfo{series}{Lecture Notes in Computer Science}} \bibinfo{volume}{1806},
  \bibinfo{publisher}{Springer Berlin / Heidelberg}, pp.
  \bibinfo{pages}{247--288}, \doi{10.1007/3-540-45594-9-9}.

\bibitemdeclare{inproceedings}{datacentricbusinessprocesses:2009}
\bibitem{datacentricbusinessprocesses:2009}
\bibinfo{author}{Alin Deutsch}, \bibinfo{author}{Richard Hull},
  \bibinfo{author}{Fabio Patrizi} \& \bibinfo{author}{Victor Vianu}
  (\bibinfo{year}{2009}): \emph{\bibinfo{title}{Automatic verification of
  data-centric business processes}}.
\newblock In: {\sl \bibinfo{booktitle}{Proceedings of the 12th International
  Conference on Database Theory}}, \bibinfo{series}{ICDT '09},
  \bibinfo{publisher}{ACM}, \bibinfo{address}{New York, NY, USA}, pp.
  \bibinfo{pages}{252--267}, \doi{10.1145/1514894.1514924}.

\bibitemdeclare{article}{logicbasedactivedatabases}
\bibitem{logicbasedactivedatabases}
\bibinfo{author}{Alvaro A.~A. Fernandes}, \bibinfo{author}{M.~Howard Williams}
  \& \bibinfo{author}{Norman~W. Paton} (\bibinfo{year}{1997}):
  \emph{\bibinfo{title}{A logic-based integration of active and deductive
  databases}}.
\newblock {\sl \bibinfo{journal}{New Gen. Comput.}}
  \bibinfo{volume}{15}(\bibinfo{number}{2}), pp. \bibinfo{pages}{205--244},
  \doi{10.1007/BF03037238}.

\bibitemdeclare{inproceedings}{places2010}
\bibitem{places2010}
\bibinfo{author}{Thomas Hildebrandt} \& \bibinfo{author}{Raghava~Rao Mukkamala}
  (\bibinfo{year}{2010}): \emph{\bibinfo{title}{Distributed Dynamic Condition
  Response Structures}}.
\newblock In: {\sl \bibinfo{booktitle}{Pre-proceedings of International
  Workshop on Programming Language Approaches to Concurrency and
  Communication-cEntric Software (PLACES 10)}}.
\newblock
  \urlprefix\url{http://www.itu.dk/people/rao/rao_files/dcrsplacescamredver.pd%
f}.

\bibitemdeclare{inproceedings}{Hull2010}
\bibitem{Hull2010}
\bibinfo{author}{Richard Hull} (\bibinfo{year}{2010}):
  \emph{\bibinfo{title}{Formal Study of Business Entities with Lifecycles: Use
  Cases, Abstract Models, and Results}}.
\newblock In: {\sl \bibinfo{booktitle}{Proceedings of 7th International
  Workshop on Web Services and Formal Methods}}, {\sl \bibinfo{series}{Lecture
  Notes in Computer Science}} \bibinfo{volume}{6551},
  \doi{10.1007/978-3-642-15618-2}.

\bibitemdeclare{article}{eventcalculusKR}
\bibitem{eventcalculusKR}
\bibinfo{author}{Robert Kowalski} (\bibinfo{year}{1992}):
  \emph{\bibinfo{title}{Database updates in the event calculus}}.
\newblock {\sl \bibinfo{journal}{J. Log. Program.}}
  \bibinfo{volume}{12}(\bibinfo{number}{1-2}), pp. \bibinfo{pages}{121--146},
  \doi{10.1016/0743-1066(92)90041-Z}.

\bibitemdeclare{inproceedings}{prohealth2008}
\bibitem{prohealth2008}
\bibinfo{author}{Karen~Marie Lyng}, \bibinfo{author}{Thomas Hildebrandt} \&
  \bibinfo{author}{Raghava~Rao Mukkamala} (\bibinfo{year}{2008}):
  \emph{\bibinfo{title}{From Paper Based Clinical Practice Guidelines to
  Declarative Workflow Management}}.
\newblock In: {\sl \bibinfo{booktitle}{Proceedings ProHealth 08 workshop}}.
\newblock
  \urlprefix\url{http://www.itu.dk/people/hilde/Papers/ProHealth08.pdf}.

\bibitemdeclare{inproceedings}{tase2010}
\bibitem{tase2010}
\bibinfo{author}{Raghava~Rao Mukkamala} \& \bibinfo{author}{Thomas Hildebrandt}
  (\bibinfo{year}{2010}): \emph{\bibinfo{title}{From Dynamic Condition Response
  Structures to {B}\"uchi Automata}}.
\newblock In: {\sl \bibinfo{booktitle}{Proceedings of 4th IEEE International
  Symposium on Theoretical Aspects of Software Engineering (TASE 2010)}},
  \doi{10.1109/TASE.2010.22}.
\newblock
  \urlprefix\url{http://www.itu.dk/people/rao/rao_files/dcrsextendedabstractTa%
se2010.pdf}.

\bibitemdeclare{inproceedings}{DDBP:2008}
\bibitem{DDBP:2008}
\bibinfo{author}{Raghava~Rao Mukkamala}, \bibinfo{author}{Thomas Hildebrandt}
  \& \bibinfo{author}{Janus~Boris T\o{}th} (\bibinfo{year}{2008}):
  \emph{\bibinfo{title}{The Resultmaker Online Consultant: From Declarative
  Workflow Management in Practice to {LTL}.}}
\newblock In: {\sl \bibinfo{booktitle}{Proceedings of DDBP}}.

\bibitemdeclare{phdthesis}{Pesic2008}
\bibitem{Pesic2008}
\bibinfo{author}{Maja Pesic} (\bibinfo{year}{2008}):
  \emph{\bibinfo{title}{Constraint-Based Workflow Management Systems: Shifting
  Control to Users}}.
\newblock Ph.D. thesis, \bibinfo{school}{Eindhoven University of Technology,
  Netherlands}.

\bibitemdeclare{inproceedings}{Senkul02alogical}
\bibitem{Senkul02alogical}
\bibinfo{author}{Pinar Senkul}, \bibinfo{author}{Michael Kifer} \&
  \bibinfo{author}{Ismail~H. Toroslu} (\bibinfo{year}{2002}):
  \emph{\bibinfo{title}{A Logical Framework for Scheduling Workflows Under
  Resource Allocation Constraints}}.
\newblock In: {\sl \bibinfo{booktitle}{In VLDB}}, pp.
  \bibinfo{pages}{694--705}.

\bibitemdeclare{inproceedings}{eventalgebra}
\bibitem{eventalgebra}
\bibinfo{author}{Munindar~P. Singh}, \bibinfo{author}{Greg Meredith},
  \bibinfo{author}{Christine Tomlinson} \& \bibinfo{author}{Paul~C. Attie}
  (\bibinfo{year}{1995}): \emph{\bibinfo{title}{An Event Algebra for Specifying
  and Scheduling Workflows}}.
\newblock In: {\sl \bibinfo{booktitle}{Proceedings of DASFAA}},
  \bibinfo{publisher}{World Scientific Press}, pp. \bibinfo{pages}{53--60}.

\bibitemdeclare{inproceedings}{PetriFair}
\bibitem{PetriFair}
\bibinfo{author}{R\"udiger Valk} \& \bibinfo{author}{Heino Carstensen}
  (\bibinfo{year}{1985}): \emph{\bibinfo{title}{Infinite Behaviour and Fairness
  in Petri Nets}}.
\newblock In \bibinfo{editor}{G.~Rozenberg}, editor: {\sl
  \bibinfo{booktitle}{Advances in Petri Nets 1984}}, {\sl
  \bibinfo{series}{Lecture Notes in Computer Science}}~\bibinfo{volume}{88},
  \bibinfo{publisher}{Springer-Verlag}, pp. \bibinfo{pages}{83--100}.

\bibitemdeclare{inproceedings}{DBLP:conf/ac/Winskel86}
\bibitem{DBLP:conf/ac/Winskel86}
\bibinfo{author}{Glynn Winskel} (\bibinfo{year}{1986}):
  \emph{\bibinfo{title}{Event Structures}}.
\newblock In \bibinfo{editor}{Wilfried Brauer}, \bibinfo{editor}{Wolfgang
  Reisig} \& \bibinfo{editor}{Grzegorz Rozenberg}, editors: {\sl
  \bibinfo{booktitle}{Advances in Petri Nets}}, {\sl \bibinfo{series}{Lecture
  Notes in Computer Science}} \bibinfo{volume}{255},
  \bibinfo{publisher}{Springer}, pp. \bibinfo{pages}{325--392}.

\bibitemdeclare{incollection}{handbookmodelsforconcurrency}
\bibitem{handbookmodelsforconcurrency}
\bibinfo{author}{Glynn Winskel} \& \bibinfo{author}{Mogens Nielsen}
  (\bibinfo{year}{1995}): \emph{\bibinfo{title}{Models for concurrency}}.
\newblock In: {\sl \bibinfo{booktitle}{Handbook of logic in computer science
  (vol. 4): semantic modelling}}, \bibinfo{publisher}{Oxford University Press},
  \bibinfo{address}{Oxford, UK}, pp. \bibinfo{pages}{1--148}.

\end{thebibliography}

\end{document}